# High temperature nanocomposites with photonic group velocity suppression of thermal emission

Andrea Mariño-López[1,2,±], Oscar Ameneiro-Prieto[1,2,±], Drew Vecchio[3,4], Xiaofei Xiao[5], Ana Sousa-Castillo[1,2,6], Nicolas Pazos-Perez[7], Stefan A. Maier[5,6,8]*, Vincenzo Giannini[9,10,11]*, Nicholas A. Kotov[3,4,12,13,14]*, Miguel A. Correa-Duarte[1,2]*, Ramon A. Alvarez-Puebla[7,15]*

[1] Department of Physical Chemistry, Universidade de Vigo, 36310 Vigo, Spain

[2] CINBIO, Universidade de Vigo, 36310 Vigo, Spain

[3] Department of Chemical Engineering, University of Michigan, Ann Arbor, MI 48109, USA.

[4] Biointerfaces Institute, University of Michigan, Ann Arbor, MI 48109, USA.

[5] Department of Physics, Imperial College London, London SW7 2AZ

[6] Chair in Hybrid Nanosystems, Nanoinstitute Munich, Faculty of Physics, Ludwig-Maximilians Universität München, Königinstrasse 10, 80539 München, Germany

[7] Department of Physical and Inorganic Chemistry, Universitat Rovira i Virgili, 43007 Tarragona, Spain.

[8] School of Physics and Astronomy, Monash University, Clayton Victoria 3800, Australia

[9] Instituto de Estructura de la Materia (IEM), Consejo Superior de Investigaciones Científicas (CSIC), Serrano 121, 28006 Madrid, Spain

[10] Technology Innovation Institute, Masdar City, Abu Dhabi 9639, United Arab Emirates

[11] Centre of Excellence ENSEMBLE3 sp. z o.o., Wolczynska 133, Warsaw, 01-919, Poland

[12] Department of Materials Science, University of Michigan, Ann Arbor, MI 48109, USA

[13] Department of Macromolecular Science and Engineering, University of Michigan, Ann Arbor, MI 48109, USA.

[14] Michigan Institute of Translational Nanotechnology (MITRAN), Ypsilanti, MI 48198, USA.

[15] ICREA, Passeig Lluis Companys 23, 08010 Barcelona, Spain.

[±] These authors contributed equally

*Correspondence to: Stefan.Maier@monash.edu (S.A.M.); v.giannini@csic.es (V.G); kotov@umich.edu (N.A.K); macorrea@uvigo.es (M.A.C.-D.); ramon.alvarez@urv.cat (R.A.A.-P.)

**ABSTRACT**: Quenching of thermal emission above 0 K is an unusual material property, essential for future energy, transportation, and space technologies. Despite the great effort invested, nearly complete quenching of thermal radiation rather than some reduction of its flux has only been achieved at low temperatures (below 373 K) and in narrow spectral windows using complex techniques suitable only for small scale objects. In this work, we present a light and flexible composite material that can suppress propagating photonic modes and, in this way, quench thermal radiation while preserving heat transfer (by thermal conduction) at a room and higher temperature below 600 K. This has been achieved by altering the local photonic density of states and consequentially the thermal properties of carbon nanotubes forming a percolating nanofiber network with a thermostable polymeric matrix.

A universal property of matter above absolute zero is thermal emission (TE), which originates from vibrations of atoms having non-zero kinetic energy. Thermal vibrations of chemical bonds give rise to fluctuating currents that produce thermal photons following a Plank spectral distribution. In recent years, we witnessed massive advances in nanostructured materials for TE engineering (*1*). Besides the fundamental interest in basic research, such studies led to marked technological progress in energy harvesting (*2*), radiative cooling (*1*), camouflage (*3, 4*), thermal concealment(*5*) and space exploration (*6*). Nonetheless, TE remains one of the primary causes of heat waste despite recent improvements in energy-efficient buildings, which can be avoided with well-designed nanomaterials (*7*).

Research on materials capable of efficient TE quenching adopts three main approaches. The first relies on the development of materials with controlled thermal emissivity, i.e. material with an absorptivity that matches the frequency requirements (*3*). In agreement with the Stefan–Boltzmann law, lower TE means low absorptivity. Thus, if we can modify the absorptivity of a material, the same is true for emissivity which can be used to match the TE of a particular object with the background, hiding objects with higher temperatures (*8-10*). The second approach focuses on the design of (meta)materials(*11*) with inhomogeneous distribution of thermal conductivity or refractive index to create thermal "cloaks" by guiding heat conduction and thermal radiation along a specific narrow direction. The third strategy involves manipulation of the surface temperature by injecting hot or cold liquid through microfluidic channels (*12*) or thermoelectric devices (*13*) under or on the surface of the object. Although successful, these approaches only work at moderate temperatures, below 373 K (100 ºC), which limits their range of applications. For instance, they are suitable for concealing objects at biological temperatures (i.e., human bodies) and, to some extent, buildings, but not vehicles. Also, most of TE quenching materials are bulky, rigid, fragile, and/or often heavy.(*14, 15*) Furthermore, their TE properties are affected by the environment, dust or moisture, while their preparation methods are complex, expensive, and limited to small areas.

Considering different alternative approaches to thermal quenching, TE strongly depends on the photonic properties of a material regardless of its absorptivity. Metallic photonic crystal will have a higher emissivity for frequencies where propagating modes are allowed and a lower one when the frequency falls within the photonic bandgap region (*16*). When the photonic local density of states (LDOS) is suppressed, so does the TE (*17*), and a well-designed photonic crystal with small LDOS can insulate, in fact, better than vacuum (*18*). The key technological and fundamental challenge is now the design of materials where suppressed LDOS is combined with other functional properties, such as strength, stiffness, toughness and lightweight, which was remains a challenge for photonic crystals especially when the material must withstand high temperatures.

The high aspect ratios of carbon nanotubes, their strong $sp^2$ carbon bonds, excellent thermal and electrical properties, and high chemical stability all contribute to making them a nearly ideal building block for multifunctional composites combining multiple extreme properties (*19*). Here, we present nanocomposites with polydimethylsiloxane (PDMS) that simultaneously display the ability to avoid heat dispersion by TE, for unprecedented high temperatures (600 K). Importantly, these TE modulation materials can be manufactured at micro and macro scales - from computer chips to residential buildings.

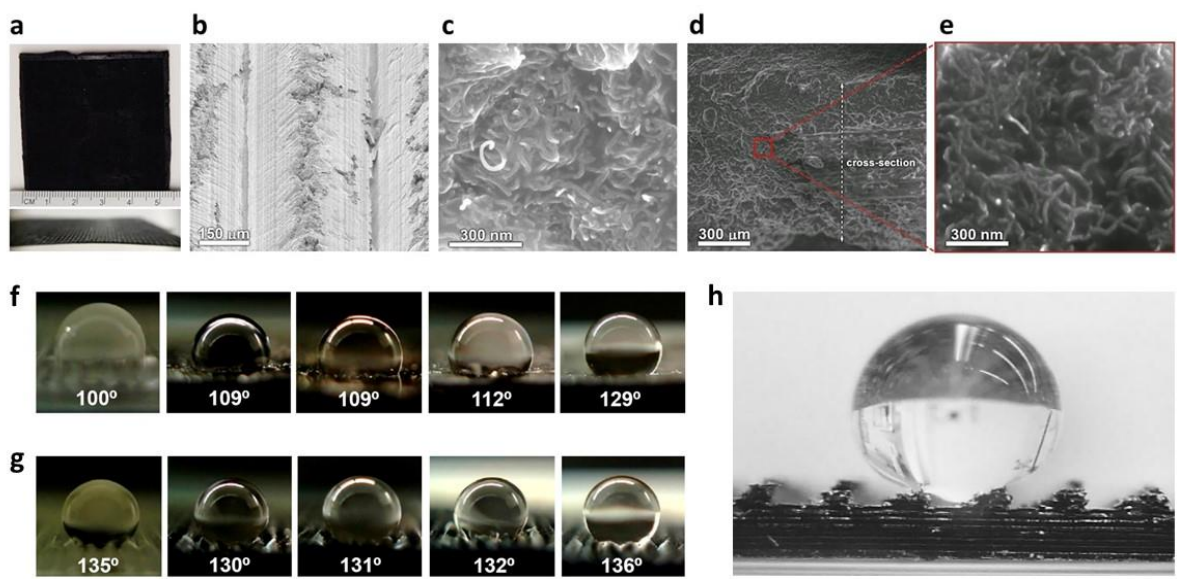

**Fig. 1.** Morphology and organization of the CNT-PDMS composite. (A) Photographs of the film as obtained after unmolding: front view (top image) and lateral view (bottom image). (B) SEM image of the top (patterned) surface of the film. (C) High-resolution SEM image of the CNTs protruding from the surface of the film. (D) Cross-sectional SEM image of the film after frozen in liquid N$_2$ and fracturing. (E) Detailed cross-sectional SEM image (red square of the previous image) of the film showing the disposition of the CNTs inside the composite. Contact angle of a droplet of water on the composite surface. (F) Contact angle of a water droplet on a smooth film as a function of the CNT concentration (from left to the right: 0, 2.5, 5, 7.5 and 10% CNT content in weight). (G) Contact angle of a water droplet on a grooved film as a function of the CNT concentration (from left to the right: 0, 2.5, 5, 7.5 and 10% CNT content in weight). (H) Detailed image of the water droplet adopting a Cassie-Baxter state on a grooved film containing 10% CNTs in weight.

Our TE modulating composites were produced by mixing multiwalled carbon nanotubes (CNTs) with average length and diameter of 1.5 μm and 9.5 nm, respectively (**Fig. 1S**), with PDMS at different ratios (from 2.5 to 10% of CNT content in weight, **Fig. S2**). To ensure a homogenous distribution of the nanotubes within the polymer matrix while preserving their physical properties, CNTs were first redispersed in a compatible solvent (hexane). An adequate mechanical force was supplied through sonication to yield good dispersion and, thus, avoid CNT aggregation in the solvent (*20*) by shielding the attractive forces between individual tubes (*21*). Then, PDMS was added, and the mixture was mechanically stirred for 12 h at 1000 rpm. This process ensures the evaporation of the solvent prior to the addition of the curing agent and a further stirring step. Finally, the resulting material was extended on a preformed surface, either smooth or patterned with grooved molds (**Fig. S3**). The desired thickness of the composite film (100-1000 μm) was obtained by applying the doctor blade method (*22*). **Fig. 1a** shows the resulting CNT-PDMS film, the dimensions of which are imposed by the size of the mold. Scanning electron microscopy imaging of the surface (**Fig. 1b**) shows a patterned surface characterized by grooves with a periodicity of 150 μm, which is consistent with the pre-existing patterning of the mold. A closer look at this surface (**Fig. 1c**) reveals how the CNT tips protrude from the polymer matrix. To ensure the chemical homogeneity of this films Raman mapping was carried out over their surface (**Fig S4**). To investigate the CNT arrangement within the bulk of the film, the material was frozen in liquid N$_2$ and fractured. Cross-sectional images of the composite material at different magnifications (**Figs. 1d** and **e**) show a "spaghetti-like" morphology of interdigitated nanofibers conducive to efficient stress transfer and high toughness (*19, 23*). Such a surface morphology imparts high hydrophobicity to the material. For a smooth surface, the progressive addition of CNTs enhances the surface hydrophobicity, as shown by the increase in water-drop contact angle from 100º to ca. 130º (**Fig. 1g**). The contact angle remains almost constant for CNT contents > 10 weight % (**Fig. 1f**) and is independent of the film thickness (**Fig. S5**). Notably, droplets adopt a minimal contact area, according to the Cassie-Baxter model (*24*), and do not penetrate the grooves as for pure PDMS film (**Fig. 1h**).

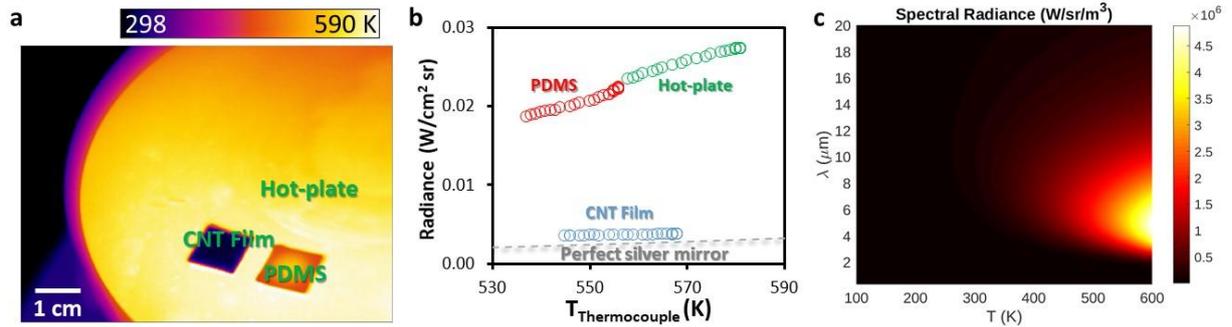

**Fig. 2.** Thermal properties of the CNT-PDMS composite film. (A) Infrared image of a hot plate equilibrated at 580 K with a CNT-PDMS composite (10% CNT content in weight) and a pure PDMS film, of 1cm of side, on top. (B) Experimental variation of the radiance of the hot plate, the composite and the pure PDMS with the temperature. The radiance of an ideal perfect silver mirror is added for comparison. (C) Thermal emission (spectral radiance) of the CNT nanocomposite obtained using eq.1 and the CNT extracted emissivity from **Fig. 2b**.

**Figure 2a** shows an IR image of a hot plate at 580 K with two square-shaped PDMS films, one constituted by the pure polymer and the other containing 10% in weight of CNT. The temperature of all surfaces was also measured with a thermocouple (**Fig S6**). Notably, while the thermocouple readouts for the hot plate and the PDMS film match those extracted from the IR measurements (580 K and 579 K for the hot plate; and 555 K and 549 K for the PDMS, respectively), in the case of the composite we observe an extraordinary divergence (566 K from the thermocouple and 311 K, from the thermal camera). Such a divergence is striking because carbon nanotube structures are known as extremely efficient broadband absorbers and, by Kirchhoff theorem, also strong emitters (*25*). In fact, they can serve as probably one of the best approximations of a black-body and amongst the darkest materials ever made (*26*).

The correlation between the irradiance provided by the IR camera and the temperature measured with the thermocouple (**Fig. 2b**) suggests that both hot plate and the PDMS film increase their photon flux with temperature, while this parameter remains almost constant for the CNT containing film. This abnormal response is very similar to the behavior of an ideal perfectly flat silver mirror (**Fig. 2b**, dotted line). From the measured thermal emission and using the Plank law (*1*):

$$B_\lambda(\lambda, T) = \frac{2hc^2}{\lambda^5} \frac{1}{e^{hc/(\lambda K_B T)} - 1} \qquad \text{(Eq. 1)},$$

we can extract the emissivity of the CNT film made, which is ~ 0.03. Such an emissivity in thermal range is outstandingly low, even assuming a perfectly flat surface, and comparable with one of the best materials available (i.e., a perfectly flat silver mirror, with an emissivity of 0.02 (*27*). This result is even more surprising considering that the CNT film is not smooth but displays roughness at the micrometre scale (**Fig. 1**). In fact, the deviation from a flat geometry commonly leads to a drastic increase the emissivity. As an example, the emissivity of silver increases almost an order of magnitude (from 0.02 to 0.1-0.14) from polished to unpolished surfaces. From the Planck law in Eq. (1) and the measured emissivity of the film (**Fig. 2b**), we can calculate the radiance per wavelength (**Fig. 2c**) showing that the TE reaches its maximum around ~5 μm (i.e., close to the roughness scale, see Supplementary Materials). This further confirms the fact that the CNT surface cannot be considered flat and that, against the experimental evidence, a high emissivity should be expected.

A possible explication of the low emissivity could be related to the high disorder of the CNT as we will detail as follows. Considering the CNT-PDMS film as a dense disordered "spaghetti-like" structure of conductive carbon nanotubes in a polymer matrix, we can infer that the system allows heat exchange by conduction from the bottom source (hot plate) to the top surface through the nanotubes, as registered by the thermocouple. At the same time, for such a nanofibrous material, a high TE would be predicted against the experimental evidence. We believe that this surprising behavior can be explained if we recognize the importance of structural heterogeneity on light propagation. Following the analysis of Schwab et al. (*28*) we focused the attention on the photonic modes in the system, $\omega_j(\mathbf{k})$, where **k** is the photon wavevector of the mode *j* and $\omega$ is the photon frequency. Thus, we can define the temperature-dependent photonic thermal conduction (*G(T)*) along the direction normal to the film as (see Supplementary Materials):

$$G(T) = \sum_j \int_{k_\perp \geq 0} \frac{d^3\mathbf{k}}{(2\pi)^3} \hbar\omega_j(\mathbf{k}) v_{g,j} \frac{\partial}{\partial T}\left(\frac{1}{e^{\hbar\omega_j(\mathbf{k})/K_BT}-1}\right)$$

(Eq. 2),

where $v_{g,j} = \partial\omega_j/\partial k_\perp$ is the group velocity of the mode *j* along the normal direction (to the film) (*13*). Accordingly, there will be only three possibilities that will allow no thermal radiation transmission to the far-field (*G = 0*). The first one requires no modes present (zero local density of states, i.e., *j = 0*). However, this event is highly improbable as conductors (as our CNTs system) have several modes allowed (*29, 30*); in fact, the local density of states is enhanced at the surface of a conductor. The second alternative demands that all supported modes are evanescent and propagating parallel to the surface, i.e., $Real(k_\perp) = 0$ (i.e., the integral has a zero domine of integration and $G(T) = 0$), but the high inhomogeneity of the fibrous nanocomposite precludes such a condition and promotes the scattering of such modes into the far-field. Thus, the only remaining option available is to consider the group velocity close to zero, $v_{g,j} = 0 \Rightarrow G(T) = 0$. In other words, the high heterogeneity of the CNT distribution in the film should lead to the formation of hot spots that localize the electromagnetic field (i.e., $v_g = 0$) and hence dramatically reduce the radiance. In support of this hypothesis, we performed a simulation of the electromagnetic field scattered by a radiating dipole located in the CNT film and analyzed the propagation of the field in such a structure.

To support this hypothesis, we built a model for TE calculations parametrized based on the SEM images (**Figure 3a**), the CNT geometry and their concentration in the film (see Supplementary Material). In this way, we built the cubic structure (1 µm edge) shown in **Figure 3b** to describe our system. Successively, we calculate light propagation in the composite using a Finite-Difference Time-Domain Method (FDTD). In FDTD methods Maxwell's equations are written in a grid of volume and time steps and solved using the central-differenced leapfrog Yee (*31*). An important point in the field calculation is to set the boundary conditions, that in this case were perfectly matched layers (PML) (*32*) (see Supplementary Materials). As our goal was to analyze the light propagation in the CNTs composite, a vertical radiating dipole was positioned at the center of the 3D structure. The wavelength of the dipolar source was 4 µm in agreement with the thermal emission peak observed in **Figure 2c** with a temperature around 600 K (and sufficient to get a good numerical convergence). In **Figure 3c**, the electric field intensity is plotted on a generic plane that is few tens of nanometer away from the dipole (and with a different saturation level in the supplementary information, see **fig S7**). As we can observe, the electromagnetic field is localized in few hotspots and its intensity decays exponentially from them. This is strong evidence supporting the theoretical argument we previously outlined: the strong disorder of CNTs does not allow light propagation but, conversely, promotes light localization, thereby drastically suppressing the thermal emission of the material. This effect depends on the CNT concentration in the film (**Fig. S8a**), where an increase in nanotube content yields a decrease in radiation transfer. For the same CNT content, film thickness does not appear to play an important role (**Fig. S8b**) being the thermal emission similar for the studied interval (film thickness = 0.2-1.2 mm). Both observations, the thermal emission dependence with CNT concentration and the irrelevance of the film thickness, agree with this light localization and zero group velocity hypothesis. In fact, as can be deduced from **Figure 3c**, no alterations of the thermal properties are expected once the film thickness is significantly larger than the typical hot-spot dimension (i.e., a few tens of nanometers).

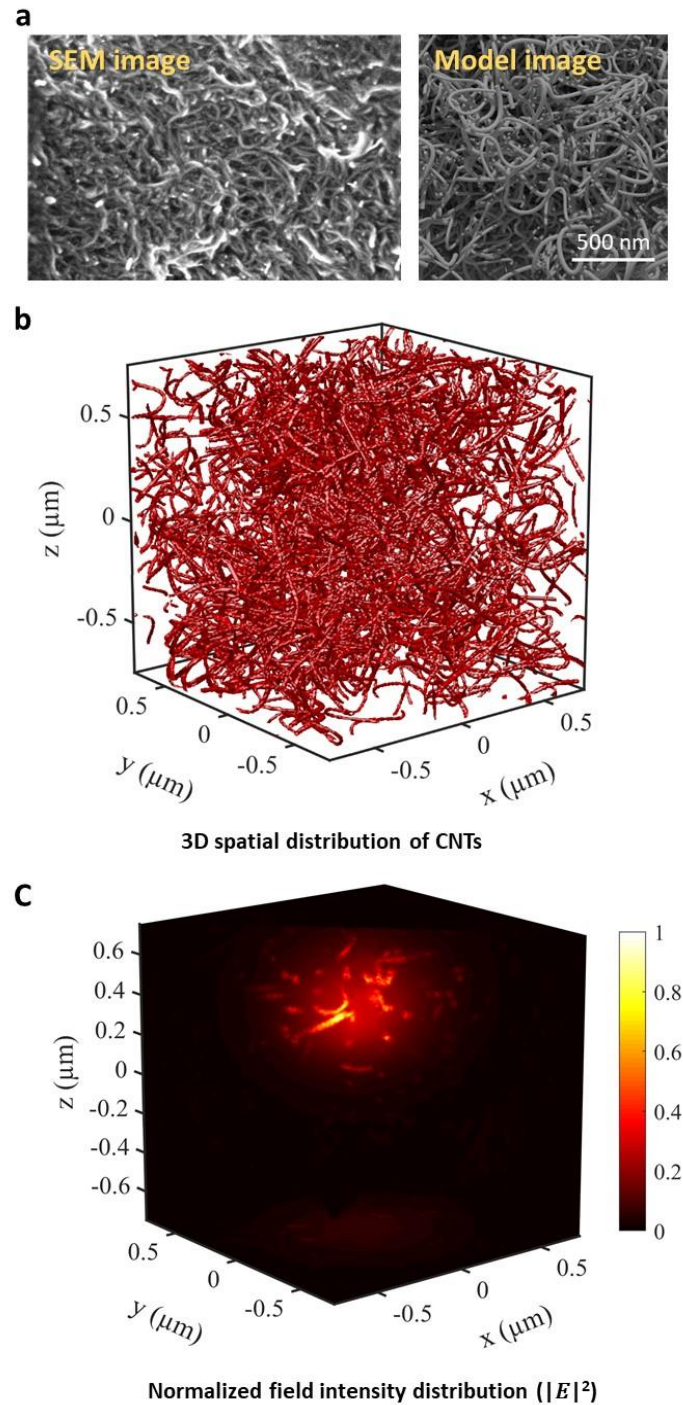

**Fig. 3.** (A) Experimental SEM of the CNT network in PDMS (10% CNT content in weight) and the model used in simulations rendered with *Blender* software. This image corresponds to a side of figure B (see SI). (B) 3D CNT structure obtained with *SolidWorks* software used in the FDTD simulations. (C) FDTD simulation of the electromagnetic field intensity of a dipole positioned at the center of the 3D CNT structure depicted (B). The normalized electric field intensity, $|\mathbf{E}|^2$, is plotted on a traversal plane a few tens nanometers apart. The creations of hot spots can be observed. The field is normalized to the maximum value obtained at 10 nm from the dipole (see also **Fig. S7**).

To further demonstrate the exceptional thermal properties of this material, we equilibrated a nanocomposite sample with a CNT content of 10% and a geometry of a 200 mm slab on a hot plate at 550 K (**Fig. 4a** and **S9,** and **movies S1** and **S2**). Then, water droplets (10 μL) were placed on the CNT-PDMS film and the bare plate surface. The droplet on the hot plate erratically moves through the surface, due to the Leidenfrost effect (*33*), until it is completely evaporated after ca. 12s. Conversely, the water droplet on the composite film retains its position while undergoing much slower evaporation (the droplet maintains the same volume for more than a minute at 550 K). Such behavior can be ascribed to two mains factors. The first one is the Cassie-Baxter state of the droplet on the superhydrophobic surface that minimizes the physical contact with the droplet surface while enabling the relaxation of the vapor layer formed at the film-droplet interphase between the grooves. This phenomenon prevents the droplet from moving. The second factor is the lack of radiation transfer through the film surface (*33*) that can be observed in **Fig. 4b** and **movies S3** and **S4**. In this experiment, the water droplet (10 μL) was placed close to a bent tin wire (100 mg) on the same CNT-PDMS film, and the temperature was progressively increased from 300 to 590 K. At 0 s (300 K), both water and tin objects are dark, which is consistent with a negligible radiation transfer. As the temperature increases (t = 60 s; 440 K), however, the hot plate brightens up and so does, to some extent, the tin wire. It is important to remind that in the experiment the thin composite film protects both objects from the plate radiation. Thus, the small radiation transfer at the top of the wire can be interpreted as the natural radiation of tin. After reaching the tin melting point (t = 110 s; 550 K), the wire begins to melt and a liquid metal droplet is fully formed at 120 s. Contrariwise, the water droplet remains stable and at low T even for a period longer than 3 min. Such different responses to the increase of the hot plate temperature can be interpreted as follows. The tin object melts because of the physical thermal conduction between the wire and the protruding CNTs in the composite film. On the other hand, in agreement with our model, the water droplet does not undergo evaporation thanks to the material's ultra-hydrophobicity and the TE suppression (that the water is warmed up by the tin droplet irradiation and not by the substrate).

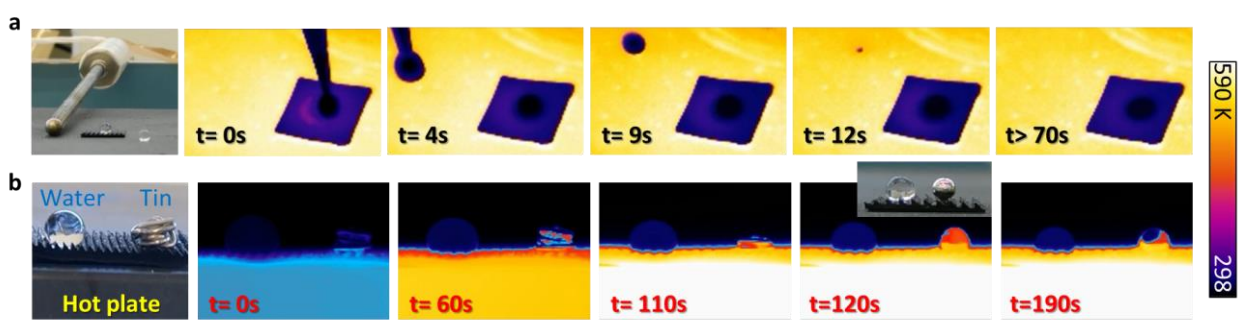

**Fig. 4.** Thermal behavior of the CNT-PDMS film in the presence of different materials. (A) Photograph and IR images for the comparison of a water droplet on the composite film vs on the bare hot-plate. (B) Photograph and infrared images for the comparison of a water droplet vs a bent tin wire on the same composite film. Optical inset shows the coexistence of both water and tin in liquid state over a surface close to 600 K. Optical and IR videos are available in the supplementary information (movies S1-4).

In summary, a material capable of TE quenching while preserving thermal conduction has been developed. The drastic reduction of propagating photonic states allows the inhibition of the radiation transfer and is related to the formation of hot spots within the peculiar architecture of interconnected nanofibers, which minimizes the photonic group velocity of thermal modes in the nanofibrous composite. This CNT-PDMS nanocomposite is thin, light, flexible and extrudable. Surface patterning of the material surface also allows for the tunability of its surface hydrophobicity which, in turn, enables unique phenomena such as the coexistence of liquid metal and liquid water on the same film. We expect that these results will pave the way for thermal management in multiple technologies and scales from $10^{-5}$ to $10^2$ m dimensions of objects.

**Acknowledgements: Funding:** This work was funded by the Ministerio de Ciencia Innovacion/AEI and the European Union Next Generation/PRTR (PDC2021-121787-I00, PID2020-120306RB-I00 and CTM2017-84050R), Xunta de Galicia/FEDER (IN607A 2018/5 and Centro Singular de Investigación de Galicia accreditation 2019–2022, ED431G 2019/06), 0712_ACUINANO_1_E, 0624_2IQBIONEURO_6_E cofounded by FEDER through the program Interreg V-A España-Portugal (POCTEP), NANOCULTURE (ERDF: 1.102.531) Interreg Atlantic Area, the European Union (European Regional Development Fund-ERDF), Generalitat de Cataluña (2017SGR883), URV (2017PFR-URV_B2-02), the Consejo Superior de Investigaciones Cientficas (INTRAMURALES 201750I039), the European Union (ERDF), the "ENSEMBLE3 - Centre of Excellence for nanophotonics, advanced materials and novel crystal growth-based technologies" project (GA No. MAB/2020/14). the EPSRC Mathematical Fundamentals of Metamaterials programme grant (EP/L024926/1) and the US Air Force Office of Scientific Research/EOARD (FA9550-17-1-0300). The NAK part of this work was supported by the NSF project "Energy- and Cost-Efficient Manufacturing Employing Nanoparticles" NSF 1463474 and Vannewar Bush DoD Fellowship to N.A.K. titled "Engineered Chiral Ceramics" ONR N000141812876. S.A.M. acknowledges the Lee Lucas chair in Physics, the Solar Technologies Go Hybrid Initiative, and the Deutsche Forschungsgemeinschaft (DFG, German Research Foundation) under Germany´s Excellence Strategy – EXC 2089/1 – 390776260. **Author contributions:** V.G., S.A.M, N.A.K., M.A.C.-D. and R.A.A.-P conceived and designed the experiments. A.M.-L., O.A.-P. and A. S.-C. prepared the materials. V.G. and X.X conducted numerical simulations. D.V. wrote the code for *StructuralGT* and performed the GT analysis of CNTs. A.M.-L., A.S.-C., N.P.-P. performed the characterization. V.G., N.A.K., M.A.C.-D. and R.A.A.-P. co-wrote the paper. M.A.C.-D. and R.A.A.-P. supervised the entire study. All authors contributed to the final version of the paper. **Competing interests:** None declared. **Data and materials availability:** All (other) data needed to evaluate the conclusions in the paper are present in the paper or the supplementary materials.

**Supplementary Materials:**

Materials and Methods

Figures S1-S9

Movies S1-S4